\documentclass[apj]{emulateapj}
\slugcomment{{\sc Accepted to ApJ:} April 1, 2016} 
\usepackage{amsmath}
\usepackage{natbib}
\usepackage{graphicx}
\usepackage{epstopdf}
\usepackage{color}
\usepackage{cancel}
\newcommand{\Fig}[1]{Figure~\ref{#1}}

\newcommand{\Rhm}{R_{\rm H,m}}

\begin{document}
\title{Orbital Stability of Multi-Planet Systems: Behavior at High Masses}
\author{Sarah J. Morrison}
\affil{Lunar and Planetary Laboratory, The University of Arizona, Tucson, AZ 85721}
\email{morrison@lpl.arizona.edu}
\and
\author{Kaitlin M. Kratter}
\affil{Steward Observatory, The University of Arizona, Tucson, AZ 85721}
\email{kkratter@email.arizona.edu}

\begin{abstract} 

In the coming years, high contrast imaging surveys are expected to reveal the characteristics of the population of wide-orbit, massive, exoplanets. To date, a handful of wide planetary mass companions are known, but only one such multi-planet system has been discovered: HR8799. For low mass planetary systems, multi-planet interactions play an important role in setting system architecture. In this paper, we explore the stability of these high mass, multi-planet systems. While empirical relationships exist that predict how system stability scales with planet spacing at low masses, we show that extrapolating to super-Jupiter masses can lead to up to an order of magnitude overestimate of stability for massive, tightly packed systems. We show that at both low and high planet masses, overlapping mean motion resonances trigger  chaotic orbital evolution, which leads to system instability. We attribute some of the difference in behavior as a function of mass to the increasing importance of second order resonances at high planet-star mass ratios. We use our tailored high mass planet results to estimate the maximum number of planets that might reside in double component debris disk systems, whose gaps may indicate the presence of massive bodies.
 
\end{abstract}


\bigskip

\section{Introduction}

With the emergence of direct imaging surveys, we are starting to probe the population of massive planets within young solar systems. Most directly imaged planets are at the upper end of the planetary mass regime with planet-star mass ratios $\mu\gtrsim10^{-3}$ and ages $\lesssim$100 Myr  \citep{Nielsen:2013, Meshkat:2015}. Several of these super-Jupiter exoplanets have been directly imaged within debris disk systems, such as in $\beta$ Pic, HR 8799, and HD 95086 \citep{Lagrange:2010, Marois:2010, Rameau:2013}. HR 8799 contains 4 directly imaged planets between its debris belts \citep{Chen:2009, Su:2009, Marois:2010}. The gap in the debris within HD 95086 is similar in size to HR 8799 \citep{Moor:2013, Su:2015}, but contains only one known planet. If planets are responsible for dynamical clearing of the gap, there may be undetected planets lurking in this system \citep{Rameau:2013, Su:2015}. 

The dynamical stability of such massive, multi-planet systems has not been explored systematically. While orbital stability has been studied extensively for lower mass planets \citep{Chambers:1996, Faber:2007, Zhou:2007, Smith:2009}, it is unclear whether these same relations should hold when the planets have substantial mass compared to their host. Most previous work has considered multi-planet stability in the context of the Hill Problem (e.g. \citealt{Gladman:1993}), even though the multi-planet results at low masses show that the metric for Hill stability, the Hill radius, does not entirely capture the behavior of systems containing more than two planets. At higher mass ratios, the approximations inherent in the Hill problem break down even in the two planet regime, suggesting that extrapolation from low mass studies may be misleading. More modest mass planets orbiting brown dwarfs and M-dwarfs may also fall beyond the reach of previous planet stability studies \citep{Bowler:2015}.

In order to place constraints on the masses and multiplicities of the planetary systems probed by direct imaging, we undertake a study of the stability of high mass ($\mu\gtrsim10^{-3}$) multi-planet systems. We report results from numerical simulations demonstrating that extrapolated trends from lower mass planetary systems fail to predict instability timescales at high masses by orders of magnitude for some planet separations. In particular, we find that at high mass ratios, there are large deviations from the monotonic trend of increasing stability with increasing intra-planet spacing. In section \ref{prevwork} we review the results from the literature. We describe our numerical integrations in section \ref{numsim} and present the results in section \ref{simres}. In section \ref{s:mmr} we compare these results to analytical estimates of instability timescales, focusing on the role of mean-motion resonance overlap across the entire planetary mass regime. We then discuss the implications for planet detection in young systems hosting debris disks in section \ref{s:obs} and summarize our findings in section \ref{s:con}.

\section{Previous Work}\label{prevwork}
We begin with a brief review of multi-planet stability.  The simplest scenario is that of two, low eccentricity planets with small planet-star mass ratios, $\mu$, in orbit about a star. In this case one can calculate analytically the Hill stability limit, which provides the minimum orbital spacing to avoid close approaches between planets \citep{Marchal:1982,Gladman:1993}. The critical spacing is often measured in units of the mutual Hill Radius of the two planets:
\begin{equation}
R_{\rm H,m}=\left(\frac{\mu_i+\mu_{i+1}}{3}\right)^{1/3}\frac{a_i+a_{i+1}}{2}
\end{equation}
where $a_i$ and $\mu_i$ are the semi-major axis mass ratio of the $i$-th planet from the star \citep{Chambers:1996}. The critical number of mutual Hill Radii is:
\begin{equation}
\Delta_{\rm Hill}=\frac{4\sqrt{3}}{2+2\sqrt{3}\left(\frac{2\mu}{3}\right)^{1/3}} \approx 2\sqrt{3}.
\end{equation} 
Note that the original \cite{Gladman:1993} definition of $\Rhm$ uses $a_i+a_{i+1} \rightarrow 2a_i$. The Hill stability limit is mass dependent in units of $\Rhm$ as defined by \citet{Chambers:1996}. This distinction is only significant for the very massive planets we consider here. Recently, \cite{Petrovich:2015} has also explored the role of high eccentricity in modifying the stability boundary.

While no firm analytic stability criteria exist for systems containing three or more planets, there are many empirical relationships derived from direct n-body integrations \citep{Chambers:1996, Faber:2007, Zhou:2007, Smith:2009}. These studies focus on the stability of $N_P\gtrsim5$ systems containing $\mu=10^{-10}-10^{-3}$ equal mass planets with equal spacings ranging from $\Delta\sim3-8$, where $\Delta$ is the number of mutual Hill Radii. They find a rough log-linear relationship between instability time (defined by orbit crossing or close planet encounter) and planet-spacing. For the mass ratios considered, multi-planet systems with $\Delta>8-10$ are typically stable over Gyr timescales. The parameter space explored to date is most relevant to low mass systems such as those discovered by \emph{Kepler}. Indeed the long-term stability predictions are borne out in multi-planet systems observed by \emph{Kepler} with typical planet separations exceeding 10$\Delta$ \citep{Lissauer:2011, Fang:2013, Fabrycky:2014, Pu:2015}.

Although numerical studies often use $\Delta$ as the distance metric for quantifying stability, previous work has revealed that stability times do not scale as $\mu^{-1/3}$ as would be predicted from two planets in the Hill problem \citep{Chambers:1996, Faber:2007, Zhou:2007, Smith:2009}. \citet{Chambers:1996} and \citet{Faber:2007} found better scaling of log time and planet spacing when separations are scaled by a factor of $\mu^{1/4}$. We denote this second spacing metric as:
\begin{equation}
\delta=\frac{a_{i+1}-a_i}{a_i \mu^{1/4}}.
\label{eq-muscale}
\end{equation}
\citet{Quillen:2011} showed analytically that the above scaling of stability with separation as a function of mass is consistent with three planets undergoing diffusion from three-body resonances until two body mean motion resonance overlap occurs. \citet{Quillen:2014} improved upon these analytic scaling relations by also accounting for the influence of nearby two-body resonances. Nevertheless, the timescales predicted by this analytic model are not as accurate as the empirical relationships fit to suites of numerical simulations.

Both the \cite{Faber:2007} and \cite{Chambers:1996} data are well fit by the functional form:
\begin{equation}\label{eq:dmu}
log(T_{\rm stab})=\mathcal{A}+\mathcal{B}{\delta}+\mathcal{C}\text{log}(\mu)
\end{equation}
with $\mathcal{A}=-8, \mathcal{B}=3.7$, and $\mathcal{C}=-1$  for the \cite{Faber:2007} results and  $\mathcal{A}=-9.11, \mathcal{B}=4.39$, and $\mathcal{C}=-1$ for \cite{Chambers:1996} (fit from \citealt{Youdin:2012}). We compare these models to our results for directly imaged planets in the Jupiter -- Super-Jupiter mass range.

\section{Numerical Simulations}\label{numsim}
In order to study the stability of high mass, multi-planet systems, we ran a series of numerical integrations to explore the impact of planet-planet spacing (initialized by $\Delta$), multiplicity ($N_P$) and planet-star mass ratio, $\mu$. We used Swifter's RADAU15 integrator for all calculations  \citep{Everhart:1985, Levison:2013}. While symplectic integrators perform well at low mass ratios, for the highest mass ratios considered here, we found that the energy conservation was $\sim10^4$ times worse than when using RADAU15. Typical integrations with the Gauss-Radau method had fractional energy changes of less than $10^{-14}$. 

We integrated three and five equal-mass planet systems separated by $\Delta=3-8$ with 
\[\mu=\{10^{-5}, 10^{-3}, 10^{-2.5}, 10^{-2}\}.\]
Because we extend our study to high mass ratios, we use Jacobi coordinates rather than central body coordinates to initialize our integrations. Using heliocentric coordinates for such high mass ratios artificially introduces modest eccentricities due to the offset of the system center of mass. All orbital parameters hereafter are reported in Jacobi coordinates. Planets are started on co-planar circular orbits about the barycenter of the enclosed masses, with constant $\Delta$ spacing. Note that the shift of the barycenter from one planet to the next does introduce small variations in the spacing for each planet as compared to a system using heliocentric coordinates. The three inner planets are set on identical orbits for $N_P=3,5$. For each combination of $\Delta$, $\mu$, and $N_P$ we simulated five systems with randomly generated initial relative orbital phases of at least 40 degrees. We only consider systems with 5 planets or fewer because higher multiplicity systems at even modest $\Delta$ would imply formation in an unphysically large and massive protoplanetary disk.

 \begin{figure}[h!]
\centering
\includegraphics[scale=0.6]{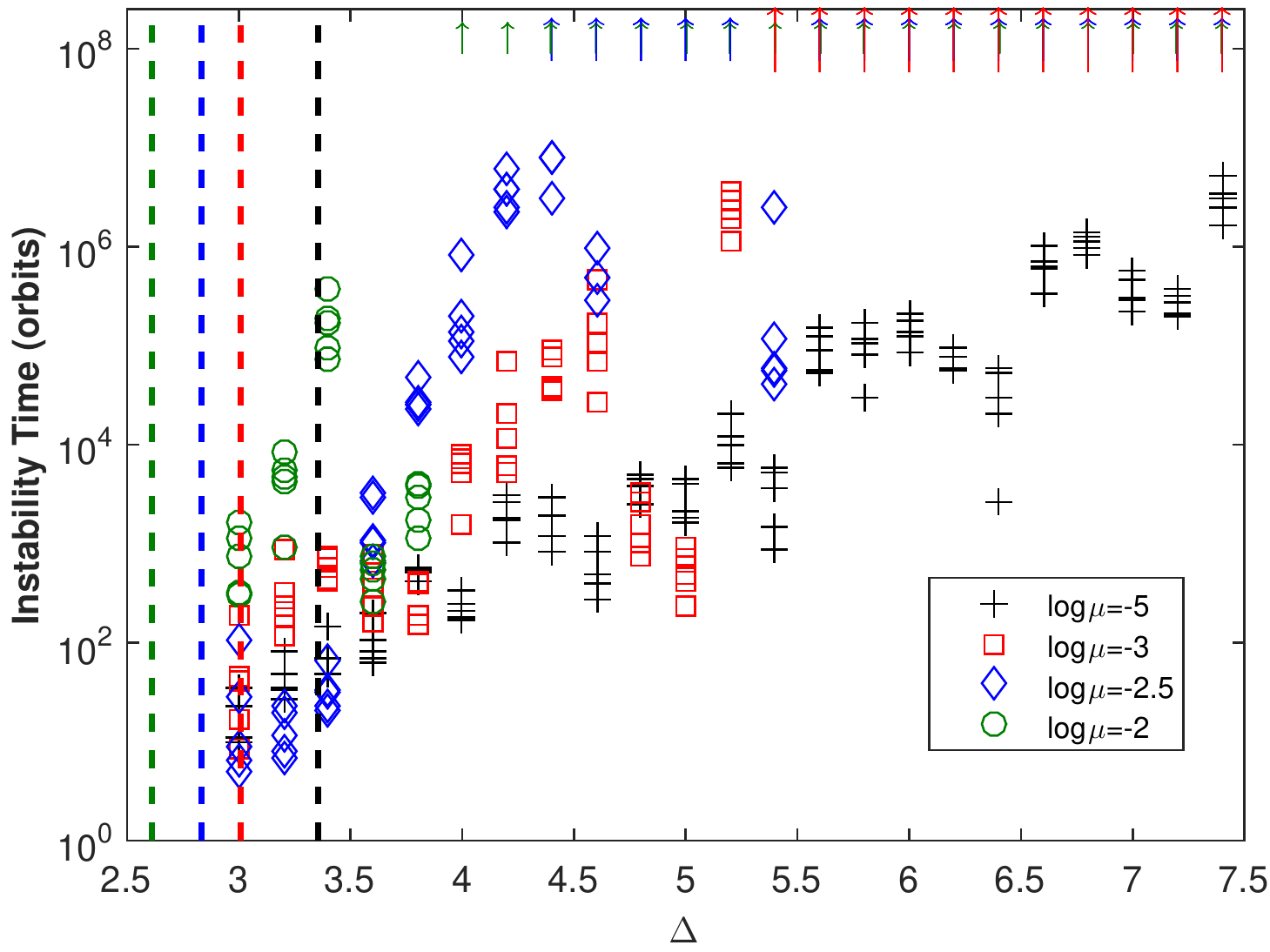}
\caption{Instability timescale (in terms of the number of innermost planet orbits) versus initial $\Delta$ for systems with 5 planets. Dashed lines delineate the Hill stability criterion for each mass. Arrows indicate lower limits to instability time based on integration durations.}
\label{fig:tvD}
\end{figure}

We integrate all systems until they become unstable or reach $10^8$ orbits of the innermost planet. Systems are deemed unstable if (1) two planets suffer a close approach (2) a planet passes beyond 1000AU from the host, or (3) a planet comes within 0.01AU of the central star.  A close approach is defined as two planets passing within a single Hill radius of each other.  We use the instantaneous heliocentric distance instead of the current osculating semi-major axis in the definition of Hill radius for calculating collisions, which better captures true close approaches for planets on high eccentricity orbits. Using orbit crossing as a criteria for system instability produces similar trends described below, albeit on shorter timescales. 

 \section{Instability Timescales for Massive Planets}\label{simres}
\begin{figure*}[h!]
\centering
\includegraphics[scale=0.75]{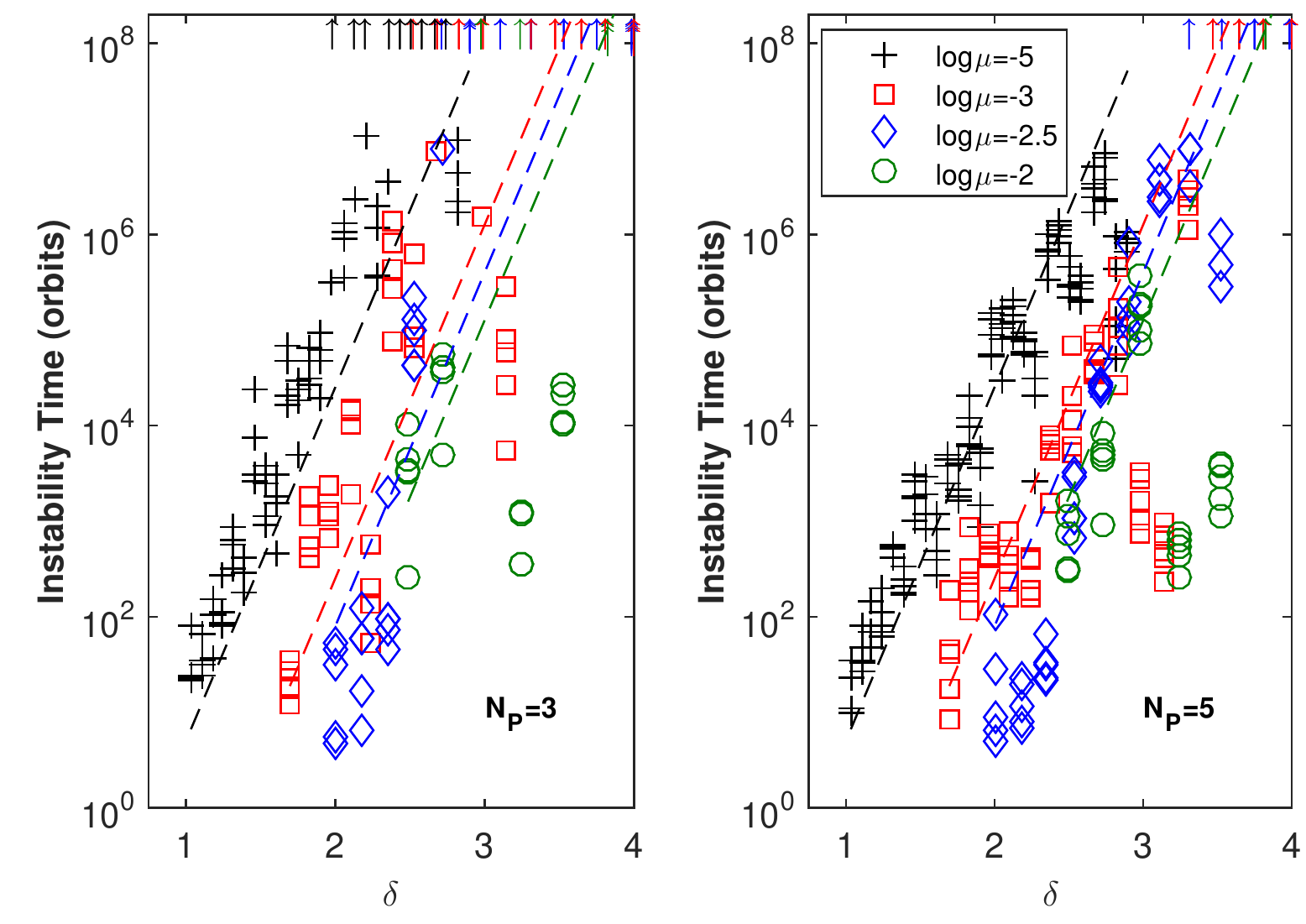}
\caption{Same as \Fig{fig:tvD} but with orbit spacings scaled by $\mu^{-1/4}$. Dashed lines are from the empirical fits for five planet systems derived from $\mu\leq10^{-3}$ numerical simulations (Eq. 2 in \citet{Faber:2007}).}
\label{fig:tvdm}
\end{figure*}

As expected, the general trend of increasing stability with increasing planet spacing holds for high $\mu$. In \Fig{fig:tvD} we show stability timescales for 5 planet systems for $10^{-5} \leq \mu \leq 10^{-2}$ as a function of $\Delta$. Here there are two noteworthy trends. First, deviations from a monotonic trend of increasing stability with increasing spacing become more pronounced at higher mass ratios. Secondly, the scaling of instability time with $\Delta$ continues to steepen at higher mass ratios:  smaller changes in $\Delta$ corresponds to a larger increases in instability time. This second trend is not entirely accounted for by adopting the $\mu^{-1/4}$ scaling from Equation \ref{eq-muscale}.  We show this scaling in \Fig{fig:tvdm} for both three and five planet systems, along with empirical fits based on the lower mass ratio data from \cite{Faber:2007}. While the $\mu^{-1/4}$ scaling adequately describes behavior of 5 planet systems at mass ratios up to $\mu\sim10^{-3}$, deviations from the scaling grow for both high mass ratios and lower multiplicity. Indeed the scatter in the 3 planet results is so high that the utility of a log-normal fit is questionable;  at $\delta <3$ for $\mu=10^{-2}$ ($\Delta =3.4$), there are systems that survive for $10^8$ years. For all mass ratios and multiplicities, the  fits become poor in proximity to mean motion resonances \citep{Zhou:2007}, however the magnitude of the discrepancy increases at high mass ratio and low multiplicity. 

Clearly for the new parameter space explored here, previous empirical fits and analytic estimates provide unreliable measures of system stability. In order to find superior empirical scaling relations for high mass ratios, it is useful to consider different metrics for measuring planet-planet spacing. In particular, measuring spacings in terms of period ratios can elucidate the driving mechanism behind system instability: mean motion resonance overlap. Although we provide empirical fits for higher mass ratios in the following sections, we urge caution when applying them to individual systems given the scatter about the fits.

\section{The Role of Mean Motion Resonance Overlap}\label{s:mmr}
The mechanism responsible for chaotic behavior on the timescales explored here is expected to be mean motion resonance (MMR) overlap (\citealt{Lecar:2001} and references therein). Neighboring first order resonances are spaced more closely as the distance toward the orbiting planet decreases, and resonance libration widths increase with planet mass. Libration widths can become large enough that planets on close and/or eccentric orbits may simultaneously feel perturbations from both resonances. The orbits become chaotic, and planets evolve on to crossing orbits or suffer close encounters after only a few hundred orbits. The extent of the region in which mean motion resonance overlap occurs has a different scaling in mass ratio than either $\Delta$ or $\delta$.

\begin{figure*}[h!]
\centering
\includegraphics[scale=0.75]{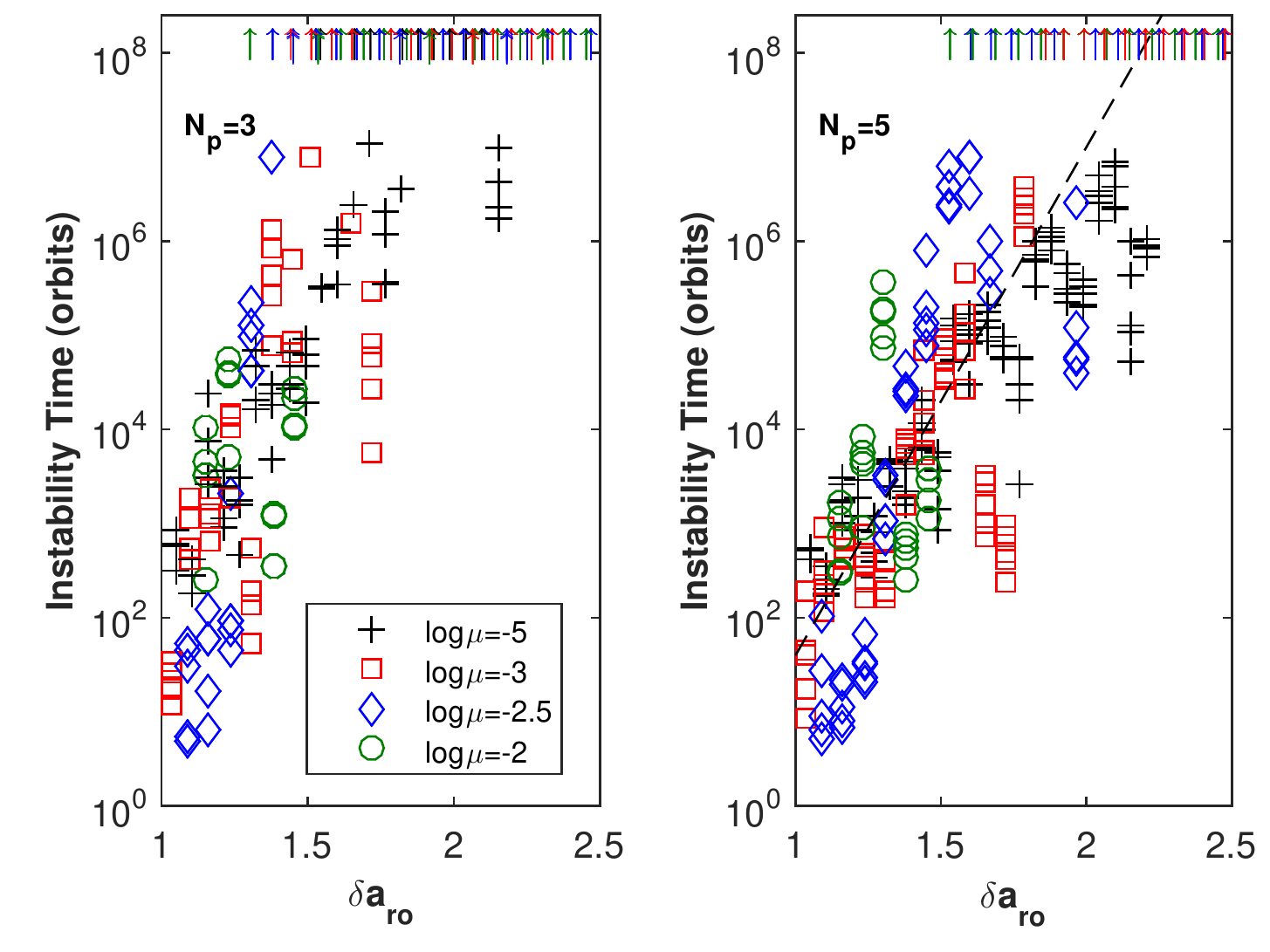}
\caption{Same as \Fig{fig:tvD} but with orbit spacings scaled by mean motion resonance overlap zone at low eccentricity (Eq.~\ref{Eq:dro}). Dashed line indicates fit to instability times shorter than the integration time for $\mu\geq10^{-3}$.}
\label{fig:tvdro}
\end{figure*}

For first order ($j+1:j$) resonances, the outermost resonance for which first order MMR overlap occurs scales as:
\begin{equation}
j_o\sim0.5\mu^{-2/7}
\end{equation}
for low eccentricity planets interacting with test particles \citep{Wisdom:1980}. The corresponding separation from a planet at which the test particle would feel both resonances is:
\begin{equation}\label{Eq:dro}
\delta a_{\rm ro}\sim 1.5\mu^{2/7}a_i,
\end{equation}
where $a_i$ is the semimajor axis. \citet{Deck:2013} showed that two massive planets follow this same relation with the substitution of $\mu=\mu_{pair},$ where $\mu_{pair}$ is the sum of the planet-star mass ratios.  
\Fig{fig:tvdro} shows instability time as a function of $\delta a_{\rm ro}$ for all 3 and 5 planet systems.  Applying this rescaling of the separations collapses most of the mass ratios down onto a single, if very broad trend. \cite{Smith:2009} also compared their numerical results with this scaling but found that when considering only a small range of mass ratios, the $\mu^{-1/4}$ scalings provided tighter fits.  

With this choice of spacing metric, neither the slope nor intercept in instability time are as mass dependent as the other metrics discussed previously. We fit a log-normal $T_{\rm stab}-\delta a_{\rm ro}$ relation for 5 planet systems with $\mu\geq10^{-3}$:
\begin{equation}
\text{log}(T_{\rm stab})=K+M\delta a_{\rm ro}
\end{equation}
where $K=-3.8$ and $M=5.4$. Typical scatter from this relation is $\sim15\%$ due in part to varying the initial relative orbital phases and the influence of individual MMRs for particular masses at a given separation. Stability timescales in the three planet case have more variation with initial phase. The fractional variation in stability timescale at a given spacing is also larger for more tightly packed systems and lower planet masses. This is consistent with the observations of \citet{Deck:2013}, who find that the transition from regular to chaotic orbits has a phase dependence close to the Hill stability limit for two planet systems. They find islands of stability for certain initial phases, although the majority of phase space is occupied by chaotic orbits. However, these islands disappear at the higher mass ratio on which we focus ($\mu>10^{-3}$) and thus the phase dependence should not be responsible for the changes in behavior we observe at high masses. Overall stability timescales are longer in systems containing three planets compared to five, resulting in $\sim$30\% steeper relations between stability timescale and adjacent period ratio and $\lesssim$10x higher stability times overall.

For the majority of our parameter space, adjacent planets on circular orbits are not initially in overlapping mean motion resonances. However, resonant libration widths are eccentricity dependent. A planet on an initially circular orbit not in resonance (or resonance overlap) can be pushed into resonance (or resonance overlap) as its eccentricity increases due to perturbations from other planets. The fractional width of the region in which 1st order mean motion resonance overlap occurs as a function of eccentricity can be estimated as:
\begin{equation}\label{Eq:joe}
{\delta a_{ro}} = 1.8 (e \mu)^{1/5}a_i
\end{equation}
\citep{Culter:2005, Mustill:2012, Deck:2013}.
Although the width of the region grows weakly with increasing eccentricity, even $e \sim 0.1$ can increase it by a factor of $\sim 2$ for $\mu\sim10^{-5}$.

Our numerical simulations suggest that a combination of eccentricity pumping and mean motion resonance overlap are responsible for orbital chaos induced instability across all mass ratios considered here. The time planets spend actually in overlapping resonances can be short, even less than our standard integration output timestep of 100 orbits for instability times several orders of magnitude longer than that. While all simulated systems surviving $\lesssim10^7$ orbits encounter one or more 1st and/or 2nd order resonances, the coarse time resolution does not always directly capture the moments at which overlap is occurring. When calculating resonance libration widths using the sum of the planet mass ratios within a pair, as has been shown to be a valid approach for estimating the extent of overlap of 1st order MMRs \citep{Deck:2013}, overlap occurs for most planet separations that survive $\lesssim10^7$ orbits.

\begin{figure*}[!]
\centering
\includegraphics[scale=0.75]{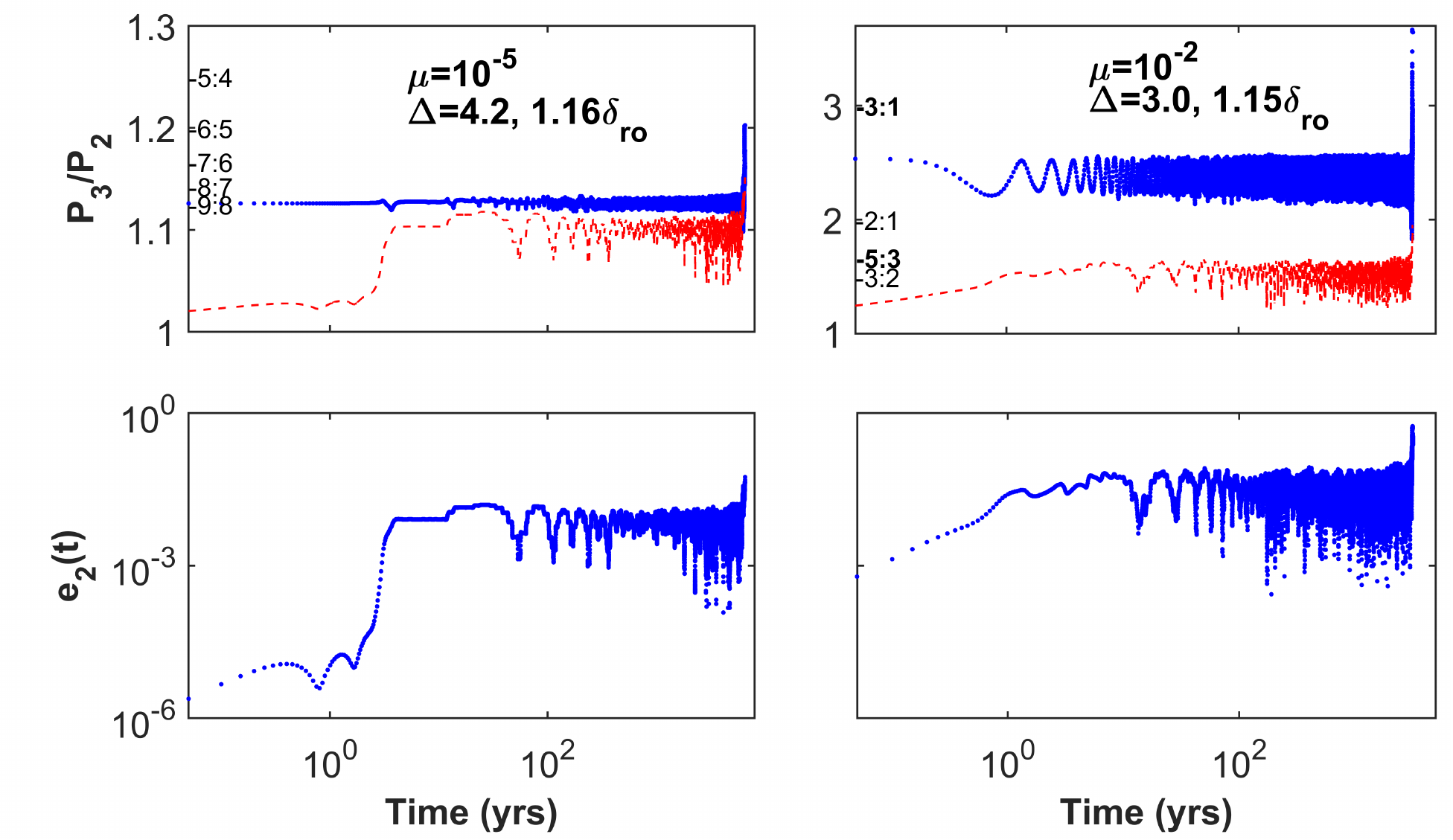} 
\caption{Time evolution of period ratio between the outer planet pair is show in blue (top plots) and the middle planet's eccentricity (bottom) for a three planet system with initial planet spacings of $\sim1.15\delta a_{\rm ro}$ for two values of $\mu$. Red lines in the two top plots show the threshold period ratio for 1st order, 2-body resonance overlap (a function of eccentricity, see Eq.~\ref{Eq:joe}). Select 1st and 2nd order (bolded) resonance period ratios are noted.}
\label{fig:tevo}
\end{figure*}

To better capture overlap, we run a subset of simulations with a higher cadence output of 0.05 orbits (choosing  those runs with instability times of $\lesssim10^4$ orbits). In all of these cases, planet undergo brief periods of overlap between 1st order MMRs, 2nd order MMRs, and/or both when using either $\mu$ or $2\mu$. As a typical example, in \Fig{fig:tevo} we show the evolution of the period ratio for a planet pair, along with the evolution of the critical period ratio at which mean motion resonance overlap occurs, including the evolving eccentricity. At both $\mu=10^{-5}$ and $\mu=10^{-2}$, the planet pairs have modest eccentricity growth until the overlap criterion is satisfied. They then undergo much wilder orbital variations before going unstable in 10s -100s of orbits. However, the eccentricity does not grow as quickly or peak at the values estimated analytically by \citet{Zhou:2007}; this is likely due to the close encounter approximation within the Hill problem breaking down at high masses for a given $\Delta$.

\begin{figure}[!]
\centering
\includegraphics[scale=0.6]{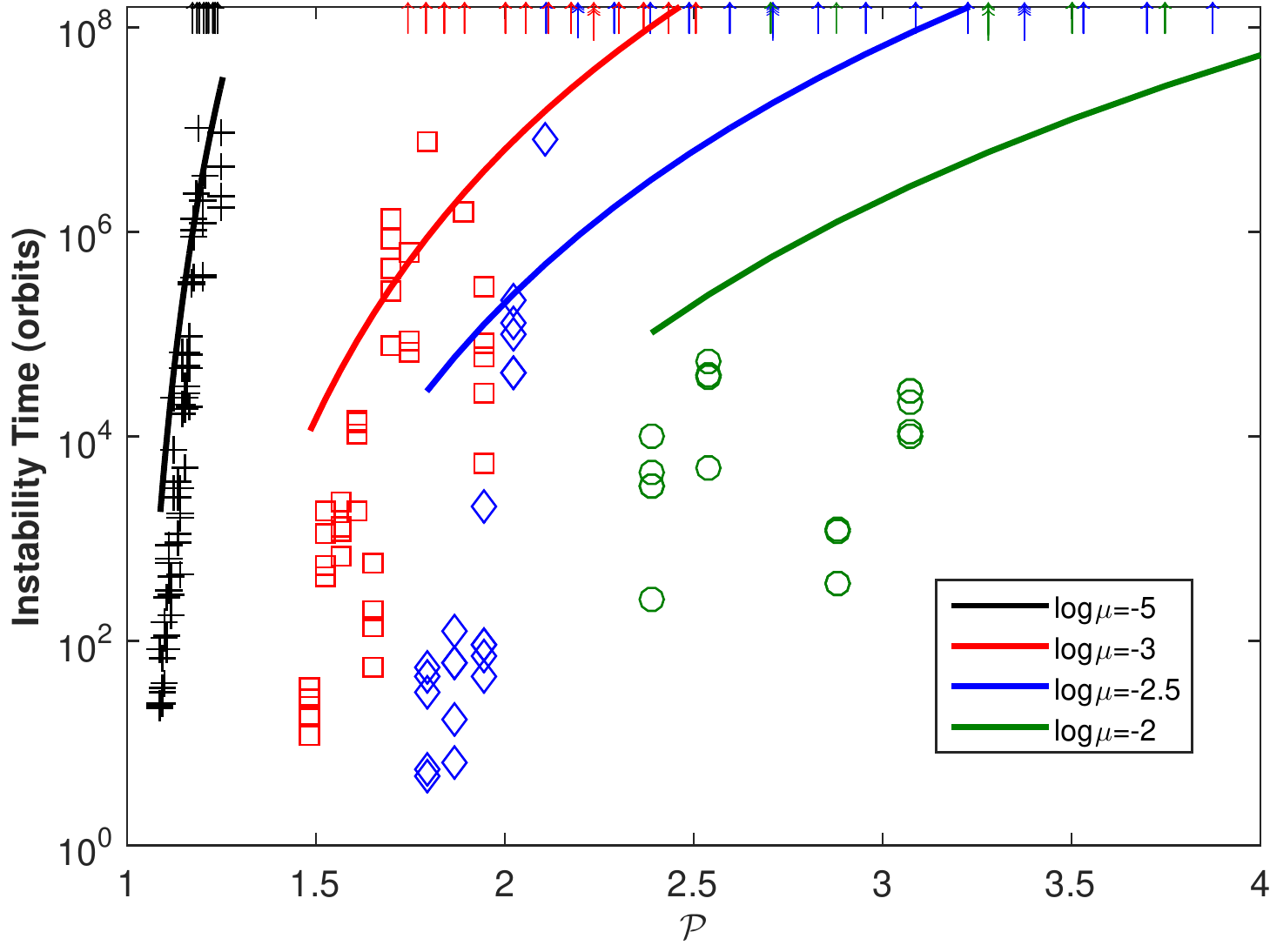}
\caption{Comparison of instability timescales from analytic estimates to numerical results for $N_P=3$ systems as a function of initial adjacent period ratio ($\mathcal{P}$). Solid lines are instability time estimates using the 3-body resonance overlap model from \citet{Quillen:2014}. Colors indicate planet mass and up arrows indicate lower limits of instability times from numerical simulations.}
\label{fig:TvPmodels}
\end{figure}

\begin{figure}[!]
\centering
\includegraphics[scale=0.6]{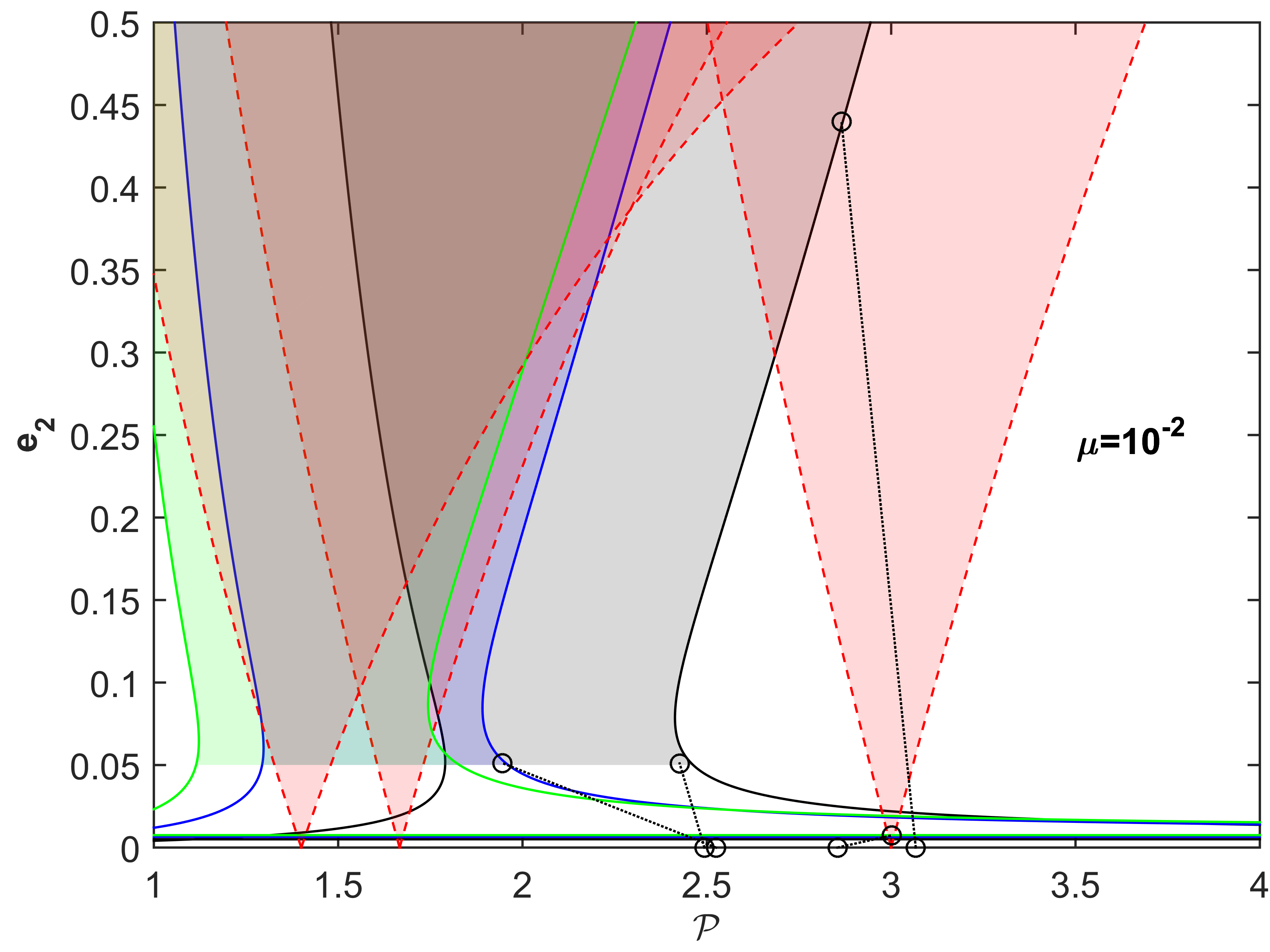} 
\caption{The three outermost first order (solid) and second order (dashed) resonance locations and widths (\citealt{Deck:2013, Bodman:2014}, and references therein) as a function of eccentricity and period ratio for $\mu=10^{-2}$. The calculation of 1st order resonance libration widths breaks down at low eccentricities, so the resonance widths are shaded only for $e>=e_{crit}$ and treated as the low eccentricity case as shown in \citet{Wisdom:1980} where $e_{crit}\sim0.5\sqrt{\mu}$. Circles show period ratio and eccentricities of planet 2 at the time of overlap or first resonance encountered connected with dashed lines to initial period ratios in $N_P=3$ simulations.}
\label{fig:reswidthsim}
\end{figure}

In \Fig{fig:reswidthsim}, we show the broad range of period ratios over which mean motion resonance overlap may occur. We overplot the initial and final period ratios and eccentricities for a handful of systems. Comparison with \Fig{fig:TvPmodels} illustrates the full parameter space we explore in terms of period ratio. Long lived systems reside near a single resonance or are widely separated beyond the 2:1 and/or the 3:1 resonances.

Despite evidence that mean motion resonance overlap is responsible for the onset of chaos across the mass range considered here, the timescale of the instability is  not well captured by existing analytic models. \Fig{fig:TvPmodels}  shows the instability time vs period ratio for three planet systems along with the analytic predictions from the \cite{Quillen:2014} 3-body+nearby 2-body 1st order resonance model.  Although the analytic models are not as accurate as the empirical fits at any mass ratio, the discrepancy gets significantly worse for high mass planetary systems, over predicting their stability. 

The instability - period ratio plot elucidates some of the physical mechanisms that may alter the behavior at high masses. In particular, one can see that while low mass ratio planets spaced at these values of $\Delta$ pile up near first order mean motion resonances, the high mass ratio systems span higher period ratios so 2nd order resonances begin to play a larger role. \Fig{fig:reswidthsim} shows that even the outermost first and second order resonances can overlap for modest eccentricities at high planet masses. The addition of a second group of resonances likely explains the deviations from monotonicity seen especially at $\mu>10^{-2.5}$ where the evolution is likely dominated by the interplay between first and second order resonances. 

The slope between instability timescale and adjacent period ratio, $\mathcal{P}$, becomes less steep with increasing mass, but approaches a constant for $\mu\gtrsim10^{-3}$ as seen in \Fig{fig:TvPmodels}. Using a similar functional form as in Eq.~\ref{eq:dmu}, we average log(instability time) of all orbital phases for each planet separation and mass to yield a least squares fit to $N_P=5$ systems of the form:
\begin{equation}
log(T_{\rm stab})=A+B\mathcal{P}+C\text{log}(\mu)
\end{equation}
for $\mu\geq10^{-3}$ yields A=-36.4, B=9.73, and C=-7.9 with an R-squared value of 0.93 when excluding systems initially near the 5:3, 2:1 and 3:1 resonances. For systems with separations close to mean motion resonances (particularly of 1st order or second order, $j+2:j$ for odd $j$), the stability timescales can be orders of magnitude shorter than the general monotonic trend. This is particularly apparent for $\mu=10^{-3}$ near the 2:1 and $\mu=10^{-2}$ near the 3:1. Adjacent planets separated by constant $\Delta$ will also have the same period ratio, $\mathcal{P}$, at any orbit distance, so dynamical effects from individual resonances should be amplified. For these high masses, $j_o\sim$ a few, so nearly all 1st order mean motion resonances are overlapping. Perturbed to sufficiently high eccentricities, high mass planets near the 5:3 and 3:1 can even experience resonance overlap with nearby 1st order resonances.

\section{Planets in ScoCen Debris Disks}\label{s:obs}
We now apply our high mass numerical stability results to gapped debris disks systems that might harbor detectable massive planets. Debris disks are excellent tracers of planet formation and evolution. There are a growing number of debris disk systems that contain two temperature components in their spectral energy distributions (SEDs). These components are interpreted as two-belt debris disks with wide gaps \citep{Kennedy:2014}. The observed dust grains in debris disks should be short lived due to Poynting-Robertson drag, and thus it is expected that they are replenished through ongoing planetesimal collisions. These collisions may be induced by dynamical perturbations from planets orbiting within the gap, which simultaneously explain the cleared central regions (\citealt{Wyatt:2008} and references therein). \citet{Faber:2007} estimated the number of low mass planets that could be responsible for such disk gaps. While these planets remain mostly beyond the reach of current imaging surveys, the high mass planets considered here are now often detectable by SPHERE, GPI, LBTI, and MagAO. Indeed a handful of planets have already been detected in debris disk systems (e.g. \citealt{Rameau:2013, Bailey:2014}).  The HR 8799 planets also reside in such a system, sharing striking similarities with our Solar System's  two belt + giant planet configuration \citep{Su:2009,Marois:2010,Moro-Martin:2010}.

Given the abundance of new instruments capable of detecting these systems,  we can use our high mass integrations to place limits on the planet configurations that would remain dynamically stable for the lifetime of these debris disk systems, extending the work of \citet{Faber:2007}.  As noted above, we emphasize that one must exercise caution given the scatter in instability timescale in this part of the parameter space. The limits we present are also derived for planets in coplanar, non-resonant configurations; resonance can both stabilize and destabilize systems, while misaligned orbital planes will likely enhance stability.

We use the sample of two component debris disks modeled by \cite{Jang-Condell:2015}. These disks are members of the $\sim$15 Myr Scorpius-Centaurus OB (ScoCen) association \citep{Pecaut:2012}. There are 44 two component disk systems characterized by \emph{Spitzer} IRS, including HD 95086 that hosts one 5$M_J$ directly imaged planet between its debris belts, and HIP 73990 with two low mass brown dwarfs outside its debris belts \citep{Rameau:2013, Hinkley:2015}.  

For each disk model presented in Table 4 of \cite{Jang-Condell:2015}\footnote{Note that there is a formatting error in the ApJ version of the paper which duplicates several entries. The first 44 lines of the table are correct, as is the arxiv version of the paper.}, we calculate the number of planets of a given mass ratio $\mu=10^{-2}, 10^{-2.5}$ and $10^{-3}$ that could fit within the two belts. We first calculate the expected stand-off distance between a planet and the inner and outer edges of the debris disk. This so-called cleared zone indicates the region from which these particles will be removed through either ejections or collisions by a planet of a given mass. We apply the cleared zone scalings from Table 1 in \cite{Morrison:2015} to calculate the inner and outermost allowed orbit for a planet in each gap. We then determine whether one or more equal mass planets spaced equally in $\Delta$ could survive in this reduced gap for the age of the system. Note that these cleared zone widths may overestimate the available dynamical space for planets because they are derived from calculations with only one planet.

 For two planet systems, we impose that planet-planet spacings must meet $\Delta\geq\Delta_{\rm Hill}$ to be stable. For $N_p=3,5$, we use the numerical integrations directly, picking the closest available simulated $\Delta$. We compare the system age to the average of the log of the instability time over all phases. By using the nearest relevant simulated system, rather than the empirical fits,  we can capture to some extent non-monotonic behavior (see Fig.~\ref{fig:tvD}).  For systems with $N_p=4$ we use the midpoint in logarithmic space between $N_p=3$ and 5 to estimate timescales. The orbital period of the innermost planet also serves to scale the instability timescales from our numerical simulations to an absolute dynamical lifetime. 

\begin{figure}[h!]
\centering
\includegraphics[scale=0.6]{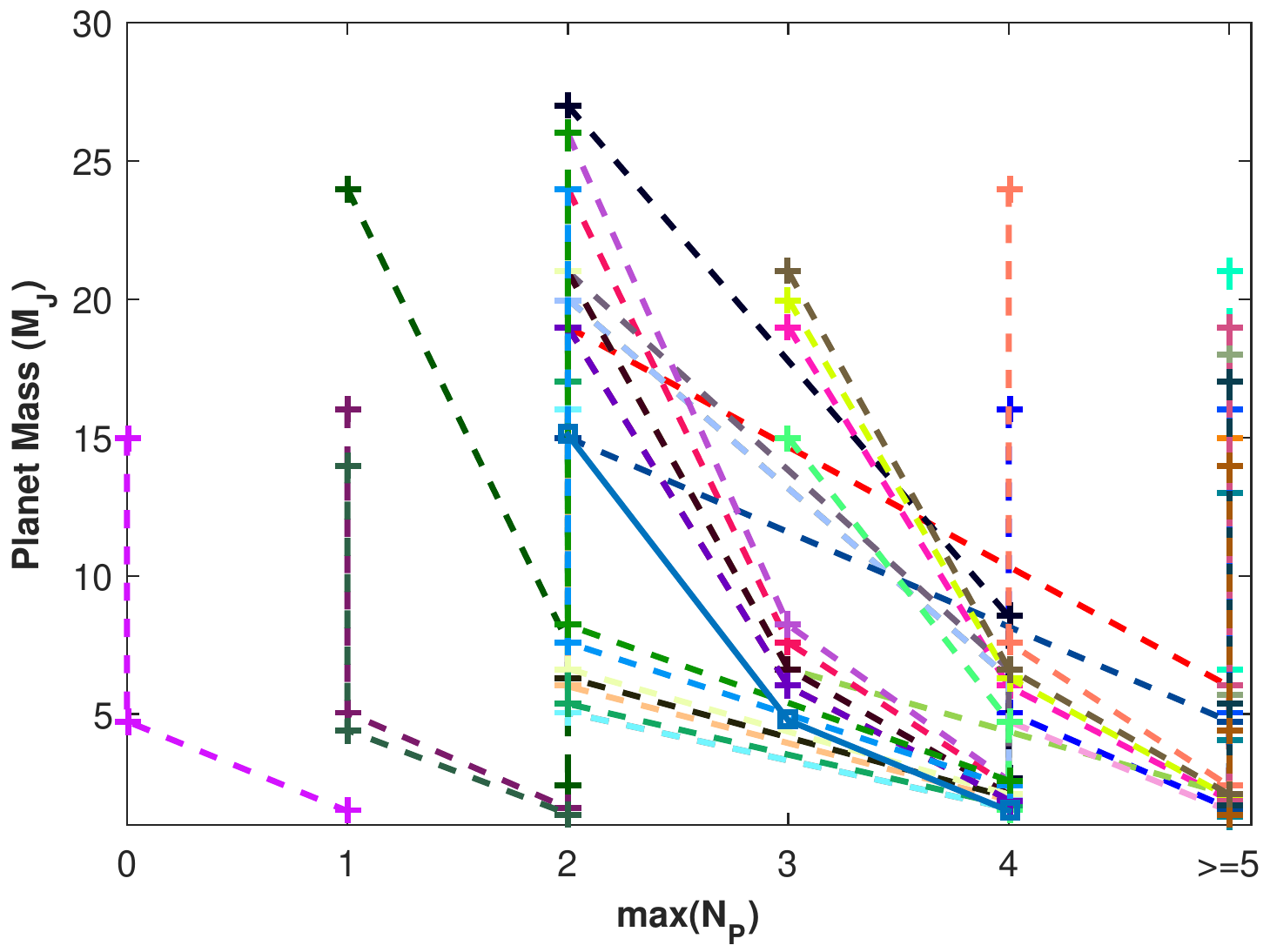}
\caption{Maximum number of planets for equal mass, equal $\Delta$ planet configurations within double debris belt systems in the ScoCen OB association based on dynamical stability. Debris belt locations are from \citet{Jang-Condell:2015}. Each color corresponds to an individual system, for which there are three points corresponding to the maximum number of planets for mass ratios of $\mu=10^{-2}, 10^{-2.5}$ or $10^{-3}$ connected by dashed lines. For comparison, predictions for HR 8799 are shown with blue squares connected by a solid line.}
\label{fig:mvNp}
\end{figure}

\Fig{fig:mvNp} shows the results of this calculation for each double debris disk system in \cite{Jang-Condell:2015}. Each system is represented by three connected points for each of the three mass ratios we tested. We use the mass of each host star to translate mass ratio into $M_{J}$ to facilitate comparison with observational detection limits. Varying the age from 11-17 Myr to reflect the age range of various subgroups of ScoCen \citep{Pecaut:2012} does not change these results significantly. The results are of course sensitive to the estimated radial distances of the debris.

Aside from the Solar System, which resides off the right hand side of the figure, HR 8799 is currently the only known multi-planet system residing between two debris belts. The allowed mass-multiplicity combinations we would predict from our simulations are shown by the thick blue squares and solid line.  We use the debris disk properties from \citep{Su:2009} and adopt a system age of 40 Myr \citep{Marois:2010, Baines:2012}. Based on this analysis, we would expect at most three $\sim5$ $M_J$ planets, as opposed to the four currently known planets \citep{Marois:2010} separated by $\Delta\sim3-4$. Astrometric fits for the HR 8799 planet orbits along with detailed dynamical modeling indicate that 3 of the planets are likely in resonance and that one of the planets may have an inclined orbit of $\sim15^o$ relative to the others \citep{Fabrycky:2010, Gozdziewski:2014, Pueyo:2015}. HR 8799 is at the threshold of stability; varying the planet orbits slightly can reduce the system's stability timescale by several orders of magnitude. \citep{Fabrycky:2010, Moore:2013, Gozdziewski:2014, Pueyo:2015}.

Based on the debris disk separations and constraints from the dynamical stability of putative planets, 21 out of 44 double debris systems in ScoCen could harbor at least one but not more than three planets over $\sim6 M_J$ for 15 Myr using the median stellar mass (1.9$M_\odot$). The other 23 systems have sufficiently widely spaced debris disks to allow for more than three planets for this mass. 28 of the 44 systems could not harbor more than two $\mu=10^{-2}$ planets/brown dwarfs.

The current estimated occurrence rates of high mass planets from direct imaging campaigns are consistent with our dynamical constraints above.
The Gemini NICI Planet-Finding Campaign estimates that fewer than 10\% of 2 $M_\odot$ stars have a 10 $M_J$ planet beyond 38 AU \citep{Nielsen:2013} and the SEEDS high-contrast imaging survey estimates giant planet occurrence around $\lesssim15-30\%$ for similar masses and separations \citep{Janson:2013}. Of the 44 systems we consider here, 11 have their outer debris belt beyond 38 AU \citep{Jang-Condell:2015}, 8 of which could also potentially host one planet $\gtrsim$ 10 $M_J$ past 38 AU. Based on the gap sizes and sensitivity limits, it is not surprising that few high mass planets have not been discovered in this sample.  However, as direct imaging campaigns from e.g. SPHERE, GPI, and LBTI push to smaller orbital separations, Super-Jupiters will be detectable in many of these systems. With a better observational survey of the parameter space, we can begin to discern whether giant multi-planet systems are responsible for debris disk gaps. We can compare detections rates with our stability analysis to determine whether such systems are similarly dynamical packed to HR 8799. If formation mechanisms typically limit the number or mass of planets in gaps (rather than dynamical stability considerations), we would expect a large discrepancy between the multiplicities calculated here and detection rates.

 \section{Conclusions}\label{s:con}

We have shown that high mass ratio planets do not obey the same scaling relationship of stability time with planet spacing as their low mass counterparts. Not only do analytic estimates for the instability times perform poorly at high masses, even empirical power law fits to the numerical data do not adequately capture the high mass ratio regime. We suggest that the deviations from monotonic trends are caused by the increasing importance of 2nd order mean motion resonances, which can overlap first order resonances at high mass ratios and eccentricities.

We also show that metrics other than mutual Hill radius may be more useful when studying high mass planets. We find that the best metric for quantifying instability as a function of spacing across all mass ratios is $\delta a_{\rm ro}$, the distance expressed as a fraction of the zone of overlapping 1st order mean motion resonances. Examining stability as a function of period ratio is also useful for identifying special locations where individual resonances come in to play. Due to the large scatter in stability timescales, detailed, system-specific dynamical modeling for high mass multi-planet systems may often be necessary, particularly to assess stability in non-planar, resonant configurations such as in the case of the four planet system HR 8799 \citep{Fabrycky:2010, Moore:2013, Gozdziewski:2014, Pueyo:2015}.

Nevertheless, numerical simulations for a large sampling of planet spacings can be valuable for placing limits on the number of high mass planets in young systems that are appealing targets for direct imaging surveys. We showed that for 44 stars hosting double debris disks in ScoCen, nearly 50\% could not harbor more than three $\sim6$ $M_J$ planets due solely to orbital stability constraints of the planetary system between the debris disks. As direct imaging observations of young planetary systems harboring debris disks continue to grow, we can begin to compare planet occurrence rates with dynamical stability constraints from the planetary system to investigate the ease of high mass planet formation and the ability to retain planets through solar system evolution.

\section{Acknowledgments}
We thank an anonymous reviewer for comments that improved this manuscript. This research was partially supported by a NASA Earth and Space Sciences Fellowship (NASA grant NNX13AO65H) and made use of the NASA Astrophysics Data System Bibliographic Services. This material is based upon work supported by the National Science Foundation under Grant No. 1228509. KMK was supported in part by NSF AST-1410174.

\bibliographystyle{apj}

\begin{thebibliography}{46}
\expandafter\ifx\csname natexlab\endcsname\relax\def\natexlab#1{#1}\fi

\bibitem[{{Bailey} {et~al.}(2014){Bailey}, {Meshkat}, {Reiter}, {Morzinski},
  {Males}, {Su}, {Hinz}, {Kenworthy}, {Stark}, {Mamajek}, {Briguglio}, {Close},
  {Follette}, {Puglisi}, {Rodigas}, {Weinberger}, \& {Xompero}}]{Bailey:2014}
{Bailey}, V., {Meshkat}, T., {Reiter}, M., {et~al.} 2014, \apjl, 780, L4

\bibitem[{{Baines} {et~al.}(2012){Baines}, {White}, {Huber}, {Jones},
  {Boyajian}, {McAlister}, {ten Brummelaar}, {Turner}, {Sturmann}, {Sturmann},
  {Goldfinger}, {Farrington}, {Riedel}, {Ireland}, {von Braun}, \&
  {Ridgway}}]{Baines:2012}
{Baines}, E.~K., {White}, R.~J., {Huber}, D., {et~al.} 2012, \apj, 761, 57

\bibitem[{{Bodman} \& {Quillen}(2014)}]{Bodman:2014}
{Bodman}, E.~H.~L., \& {Quillen}, A.~C. 2014, \mnras, 440, 1753

\bibitem[{{Bowler} {et~al.}(2015){Bowler}, {Liu}, {Shkolnik}, \&
  {Tamura}}]{Bowler:2015}
{Bowler}, B.~P., {Liu}, M.~C., {Shkolnik}, E.~L., \& {Tamura}, M. 2015, \apjs,
  216, 7

\bibitem[{{Chambers} {et~al.}(1996){Chambers}, {Wetherill}, \&
  {Boss}}]{Chambers:1996}
{Chambers}, J.~E., {Wetherill}, G.~W., \& {Boss}, A.~P. 1996, \icarus, 119, 261

\bibitem[{{Chen} {et~al.}(2009){Chen}, {Sheehan}, {Watson}, {Manoj}, \&
  {Najita}}]{Chen:2009}
{Chen}, C.~H., {Sheehan}, P., {Watson}, D.~M., {Manoj}, P., \& {Najita}, J.~R.
  2009, \apj, 701, 1367

\bibitem[{{Culter}(2005)}]{Culter:2005}
{Culter}, C. 2005, {The 1/5 Law for Chaos in the Three-Body Problem at Moderate
  Eccentricity}, UC Berkeley Honors Thesis in Physics,
  http://w.astro.berkeley.edu/~echiang/students/thesis.pdf

\bibitem[{{Deck} {et~al.}(2013){Deck}, {Payne}, \& {Holman}}]{Deck:2013}
{Deck}, K.~M., {Payne}, M., \& {Holman}, M.~J. 2013, \apj, 774, 129

\bibitem[{{Everhart}(1985)}]{Everhart:1985}
{Everhart}, E. 1985, in Dynamics of Comets: Their Origin and Evolution,
  Proceedings of IAU Colloq. 83, held in Rome, Italy, June 11-15, 1984. Edited
  by Andrea Carusi and Giovanni B. Valsecchi. Dordrecht: Reidel, Astrophysics
  and Space Science Library. Volume 115, 1985, p.185, ed. A.~{Carusi} \& G.~B.
  {Valsecchi}, 185

\bibitem[{{Faber} \& {Quillen}(2007)}]{Faber:2007}
{Faber}, P., \& {Quillen}, A.~C. 2007, \mnras, 382, 1823

\bibitem[{{Fabrycky} \& {Murray-Clay}(2010)}]{Fabrycky:2010}
{Fabrycky}, D.~C., \& {Murray-Clay}, R.~A. 2010, \apj, 710, 1408

\bibitem[{{Fabrycky} {et~al.}(2014){Fabrycky}, {Lissauer}, {Ragozzine}, {Rowe},
  {Steffen}, {Agol}, {Barclay}, {Batalha}, {Borucki}, {Ciardi}, {Ford},
  {Gautier}, {Geary}, {Holman}, {Jenkins}, {Li}, {Morehead}, {Morris},
  {Shporer}, {Smith}, {Still}, \& {Van Cleve}}]{Fabrycky:2014}
{Fabrycky}, D.~C., {Lissauer}, J.~J., {Ragozzine}, D., {et~al.} 2014, \apj,
  790, 146

\bibitem[{{Fang} \& {Margot}(2013)}]{Fang:2013}
{Fang}, J., \& {Margot}, J.-L. 2013, \apj, 767, 115

\bibitem[{{Gladman}(1993)}]{Gladman:1993}
{Gladman}, B. 1993, \icarus, 106, 247

\bibitem[{{Go{\'z}dziewski} \& {Migaszewski}(2014)}]{Gozdziewski:2014}
{Go{\'z}dziewski}, K., \& {Migaszewski}, C. 2014, \mnras, 440, 3140

\bibitem[{{Hinkley} {et~al.}(2015){Hinkley}, {Kraus}, {Ireland}, {Cheetham},
  {Carpenter}, {Tuthill}, {Lacour}, {Evans}, \& {Haubois}}]{Hinkley:2015}
{Hinkley}, S., {Kraus}, A.~L., {Ireland}, M.~J., {et~al.} 2015, \apjl, 806, L9

\bibitem[{{Jang-Condell} {et~al.}(2015){Jang-Condell}, {Chen}, {Mittal},
  {Manoj}, {Watson}, {Lisse}, {Nesvold}, \& {Kuchner}}]{Jang-Condell:2015}
{Jang-Condell}, H., {Chen}, C.~H., {Mittal}, T., {et~al.} 2015, \apj, 808, 167

\bibitem[{{Janson} {et~al.}(2013){Janson}, {Brandt}, {Moro-Mart{\'{\i}}n},
  {Usuda}, {Thalmann}, {Carson}, {Goto}, {Currie}, {McElwain}, {Itoh},
  {Fukagawa}, {Crepp}, {Kuzuhara}, {Hashimoto}, {Kudo}, {Kusakabe}, {Abe},
  {Brandner}, {Egner}, {Feldt}, {Grady}, {Guyon}, {Hayano}, {Hayashi},
  {Hayashi}, {Henning}, {Hodapp}, {Ishii}, {Iye}, {Kandori}, {Knapp}, {Kwon},
  {Matsuo}, {Miyama}, {Morino}, {Nishimura}, {Pyo}, {Serabyn}, {Suenaga},
  {Suto}, {Suzuki}, {Takahashi}, {Takami}, {Takato}, {Terada}, {Tomono},
  {Turner}, {Watanabe}, {Wisniewski}, {Yamada}, {Takami}, \&
  {Tamura}}]{Janson:2013}
{Janson}, M., {Brandt}, T.~D., {Moro-Mart{\'{\i}}n}, A., {et~al.} 2013, \apj,
  773, 73

\bibitem[{{Kennedy} \& {Wyatt}(2014)}]{Kennedy:2014}
{Kennedy}, G.~M., \& {Wyatt}, M.~C. 2014, \mnras, 444, 3164

\bibitem[{{Lagrange} {et~al.}(2010){Lagrange}, {Bonnefoy}, {Chauvin}, {Apai},
  {Ehrenreich}, {Boccaletti}, {Gratadour}, {Rouan}, {Mouillet}, {Lacour}, \&
  {Kasper}}]{Lagrange:2010}
{Lagrange}, A.-M., {Bonnefoy}, M., {Chauvin}, G., {et~al.} 2010, Science, 329,
  57

\bibitem[{{Lecar} {et~al.}(2001){Lecar}, {Franklin}, {Holman}, \&
  {Murray}}]{Lecar:2001}
{Lecar}, M., {Franklin}, F.~A., {Holman}, M.~J., \& {Murray}, N.~J. 2001,
  \araa, 39, 581

\bibitem[{{Levison} \& {Duncan}(2013)}]{Levison:2013}
{Levison}, H.~F., \& {Duncan}, M.~J. 2013, {SWIFT: A solar system integration
  software package}, Astrophysics Source Code Library

\bibitem[{{Lissauer} {et~al.}(2011){Lissauer}, {Ragozzine}, {Fabrycky},
  {Steffen}, {Ford}, {Jenkins}, {Shporer}, {Holman}, {Rowe}, {Quintana},
  {Batalha}, {Borucki}, {Bryson}, {Caldwell}, {Carter}, {Ciardi}, {Dunham},
  {Fortney}, {Gautier}, {Howell}, {Koch}, {Latham}, {Marcy}, {Morehead}, \&
  {Sasselov}}]{Lissauer:2011}
{Lissauer}, J.~J., {Ragozzine}, D., {Fabrycky}, D.~C., {et~al.} 2011, \apjs,
  197, 8

\bibitem[{{Marchal} \& {Bozis}(1982)}]{Marchal:1982}
{Marchal}, C., \& {Bozis}, G. 1982, Celestial Mechanics, 26, 311

\bibitem[{{Marois} {et~al.}(2010){Marois}, {Zuckerman}, {Konopacky},
  {Macintosh}, \& {Barman}}]{Marois:2010}
{Marois}, C., {Zuckerman}, B., {Konopacky}, Q.~M., {Macintosh}, B., \&
  {Barman}, T. 2010, \nat, 468, 1080

\bibitem[{{Meshkat} {et~al.}(2015){Meshkat}, {Kenworthy}, {Reggiani}, {Quanz},
  {Mamajek}, \& {Meyer}}]{Meshkat:2015}
{Meshkat}, T., {Kenworthy}, M.~A., {Reggiani}, M., {et~al.} 2015, \mnras, 453,
  2533

\bibitem[{{Mo{\'o}r} {et~al.}(2013){Mo{\'o}r}, {{\'A}brah{\'a}m},
  {K{\'o}sp{\'a}l}, {Szab{\'o}}, {Apai}, {Balog}, {Csengeri}, {Grady},
  {Henning}, {Juh{\'a}sz}, {Kiss}, {Pascucci}, {Szul{\'a}gyi}, \&
  {Vavrek}}]{Moor:2013}
{Mo{\'o}r}, A., {{\'A}brah{\'a}m}, P., {K{\'o}sp{\'a}l}, {\'A}., {et~al.} 2013,
  \apjl, 775, L51

\bibitem[{{Moore} \& {Quillen}(2013)}]{Moore:2013}
{Moore}, A., \& {Quillen}, A.~C. 2013, \mnras, 430, 320

\bibitem[{{Moro-Mart{\'{\i}}n} {et~al.}(2010){Moro-Mart{\'{\i}}n}, {Malhotra},
  {Bryden}, {Rieke}, {Su}, {Beichman}, \& {Lawler}}]{Moro-Martin:2010}
{Moro-Mart{\'{\i}}n}, A., {Malhotra}, R., {Bryden}, G., {et~al.} 2010, \apj,
  717, 1123

\bibitem[{{Morrison} \& {Malhotra}(2015)}]{Morrison:2015}
{Morrison}, S., \& {Malhotra}, R. 2015, \apj, 799, 41

\bibitem[{{Mustill} \& {Wyatt}(2012)}]{Mustill:2012}
{Mustill}, A.~J., \& {Wyatt}, M.~C. 2012, \mnras, 419, 3074

\bibitem[{{Nielsen} {et~al.}(2013){Nielsen}, {Liu}, {Wahhaj}, {Biller},
  {Hayward}, {Close}, {Males}, {Skemer}, {Chun}, {Ftaclas}, {Alencar},
  {Artymowicz}, {Boss}, {Clarke}, {de Gouveia Dal Pino}, {Gregorio-Hetem},
  {Hartung}, {Ida}, {Kuchner}, {Lin}, {Reid}, {Shkolnik}, {Tecza}, {Thatte}, \&
  {Toomey}}]{Nielsen:2013}
{Nielsen}, E.~L., {Liu}, M.~C., {Wahhaj}, Z., {et~al.} 2013, \apj, 776, 4

\bibitem[{{Pecaut} {et~al.}(2012){Pecaut}, {Mamajek}, \& {Bubar}}]{Pecaut:2012}
{Pecaut}, M.~J., {Mamajek}, E.~E., \& {Bubar}, E.~J. 2012, \apj, 746, 154

\bibitem[{{Petrovich}(2015)}]{Petrovich:2015}
{Petrovich}, C. 2015, \apj, 808, 120

\bibitem[{{Pu} \& {Wu}(2015)}]{Pu:2015}
{Pu}, B., \& {Wu}, Y. 2015, \apj, 807, 44

\bibitem[{{Pueyo} {et~al.}(2015){Pueyo}, {Soummer}, {Hoffmann}, {Oppenheimer},
  {Graham}, {Zimmerman}, {Zhai}, {Wallace}, {Vescelus}, {Veicht}, {Vasisht},
  {Truong}, {Sivaramakrishnan}, {Shao}, {Roberts}, {Roberts}, {Rice}, {Parry},
  {Nilsson}, {Lockhart}, {Ligon}, {King}, {Hinkley}, {Hillenbrand}, {Hale},
  {Dekany}, {Crepp}, {Cady}, {Burruss}, {Brenner}, {Beichman}, \&
  {Baranec}}]{Pueyo:2015}
{Pueyo}, L., {Soummer}, R., {Hoffmann}, J., {et~al.} 2015, \apj, 803, 31

\bibitem[{{Quillen}(2011)}]{Quillen:2011}
{Quillen}, A.~C. 2011, \mnras, 418, 1043

\bibitem[{{Quillen} \& {French}(2014)}]{Quillen:2014}
{Quillen}, A.~C., \& {French}, R.~S. 2014, \mnras, 445, 3959

\bibitem[{{Rameau} {et~al.}(2013){Rameau}, {Chauvin}, {Lagrange}, {Boccaletti},
  {Quanz}, {Bonnefoy}, {Girard}, {Delorme}, {Desidera}, {Klahr}, {Mordasini},
  {Dumas}, \& {Bonavita}}]{Rameau:2013}
{Rameau}, J., {Chauvin}, G., {Lagrange}, A.-M., {et~al.} 2013, \apjl, 772, L15

\bibitem[{{Smith} \& {Lissauer}(2009)}]{Smith:2009}
{Smith}, A.~W., \& {Lissauer}, J.~J. 2009, \icarus, 201, 381

\bibitem[{{Su} {et~al.}(2015){Su}, {Morrison}, {Malhotra}, {Smith}, {Balog}, \&
  {Rieke}}]{Su:2015}
{Su}, K.~Y.~L., {Morrison}, S., {Malhotra}, R., {et~al.} 2015, \apj, 799, 146

\bibitem[{{Su} {et~al.}(2009){Su}, {Rieke}, {Stapelfeldt}, {Malhotra},
  {Bryden}, {Smith}, {Misselt}, {Moro-Martin}, \& {Williams}}]{Su:2009}
{Su}, K.~Y.~L., {Rieke}, G.~H., {Stapelfeldt}, K.~R., {et~al.} 2009, \apj, 705,
  314

\bibitem[{{Wisdom}(1980)}]{Wisdom:1980}
{Wisdom}, J. 1980, \aj, 85, 1122

\bibitem[{{Wyatt}(2008)}]{Wyatt:2008}
{Wyatt}, M.~C. 2008, \araa, 46, 339

\bibitem[{{Youdin} {et~al.}(2012){Youdin}, {Kratter}, \&
  {Kenyon}}]{Youdin:2012}
{Youdin}, A.~N., {Kratter}, K.~M., \& {Kenyon}, S.~J. 2012, \apj, 755, 17

\bibitem[{{Zhou} {et~al.}(2007){Zhou}, {Lin}, \& {Sun}}]{Zhou:2007}
{Zhou}, J.-L., {Lin}, D.~N.~C., \& {Sun}, Y.-S. 2007, \apj, 666, 423

\end{thebibliography}

\end{document}